\documentstyle[psfig]{mn}

\def\ltsima{$\; \buildrel < \over \sim \;$}
\def\lsim{\lower.5ex\hbox{\ltsima}}
\def\gtsima{$\; \buildrel > \over \sim \;$}
\def\gsim{\lower.5ex\hbox{\gtsima}}

\begin{document}

\title[Constraints on the emission mechanisms of gamma--ray bursts]
{Constraints on the emission mechanisms of gamma-ray bursts}
\author[Ghisellini, Celotti and Lazzati]
{Gabriele Ghisellini$^1$, Annalisa Celotti$^2$ and Davide Lazzati$^{1,3}$,\\ 
$^1$ Osservatorio Astronomico di Brera, Via Bianchi 46, I--23807
Merate (Lc), Italy \\
$^2$ SISSA, Via Beirut 2--4, I--34014 Trieste, Italy \\
$^3$ Dipartimento di Fisica, Universit\`a degli Studi di Milano,
Via Celoria 16, I--20133 Milano, Italy \\
E--mail: {\tt gabriele@merate.mi.astro.it}, {\tt celotti@sissa.it},
        {\tt lazzati@merate.mi.astro.it}
}


\maketitle

\begin{abstract}
If the emission of gamma--ray bursts were due to the synchrotron process
in the standard internal shock scenario, then the typical observed
spectrum should have a slope $F_\nu \propto \nu^{-1/2}$, which
strongly conflicts with the much harder spectra observed.  
This directly follows from the cooling time being
much shorter than the dynamical time.
Particle re--acceleration, deviations from equipartition, fastly
changing magnetic fields and adiabatic losses are found to be inadequate to 
account for this discrepancy.
We also find that in the internal shock scenario the relativistic inverse 
Compton scattering is always as important as the synchrotron process,
and faces the same problems.
This indicates that the burst emission is not produced by relativistic 
electrons emitting synchrotron and inverse Compton radiation.
\end{abstract}


\section{Introduction}

Since the observational breakthrough by $Beppo$SAX (Costa et al. 1997;
van Paradijs et al. 1997) the physics of gamma--ray bursts (GRB) 
has started to be disclosed.
The huge energy and power releases required by their cosmological
distances support the fireball scenario (Cavallo \& Rees 1978; Rees
\& M\'esz\'aros 1992; M\'esz\'aros \& Rees 1993), whose evolution and
behavior is (unfortunately) largely independent of their origin.

We do not know yet in any detail how the GRB event is related to the
afterglow emission, but in the most accepted picture of formation of
and emission from internal/external shocks (Rees \& M\'esz\'aros 1992;
Rees \& M\'esz\'aros 1994; Sari \& Piran 1997), the former is due to
collisions of pairs of relativistic shells (internal shocks), while
the latter is generated by the collisionless shocks produced by shells
interacting with the interstellar medium (external shocks).
The short spikes ($t_{\rm var}\sim$10 ms) observed in the high energy
light curves suggest that shell--shell collisions occur at distances
$R\simeq 10^{12}$--$10^{13}$ cm from the central source, involving 
plasma moving with bulk Lorenz factor $\Gamma \ge 100$.
The fireball starts to be decelerated by the interstellar medium
further out, at a distance which depends on the density of this material.

The main radiation mechanism assumed to be responsible for both
the burst event and the afterglow is synchrotron (Rees \&
M\'esz\'aros 1994; Sari, Narayan \& Piran 1996; Sari \& Piran 1997;
Panaitescu \& M\'esz\'aros 1998 -- see however Thompson 1994; Liang 1997; 
Ghisellini \& Celotti 1999; Celotti \& Ghisellini 1999; Stern 1999).
This requires acceleration of electrons up to ultra--relativistic
energies and the presence of a significant magnetic field.
Evidence supporting that the afterglow emission is 
due to the synchrotron process include the power law decay in time
of the afterglow flux (for reviews see Piran 1999; M\'esz\'aros 1999) 
and the recently detected linear polarization in GRB~990510
(Covino et al. 1999; Wijers et al. 1999), but the only piece of circumstantial
evidence in favor of a synchrotron origin of the burst radiation
comes from the predicted frequency of the peak of the burst spectrum.
Indeed, it is remarkable that the simple assumption of equipartition among
protons, electrons and magnetic field energy densities leads --
in the internal shock scenario (ISS) -- to a typical emission 
frequency in agreement with observations.

However, the very same ISS inevitably predicts very fast
radiative cooling of the emitting particles.
In this Letter we point out that this implies an emitted spectrum
much steeper than observed.
Although other authors have already pointed out that, in the presence of
radiative losses, the predicted spectrum is steep (Cohen et al. 1997;
Sari, Piran \& Narayan 1998; Chiang 1999), here possible alternatives 
to avoid this conclusion, in the context of the ISS, are discussed, but 
found inadequate to account for the discrepancy.
In addition, we examine the role of the relativistic inverse Compton
process in the ISS, which results to be as important as the
synchrotron one.
Therefore in this scenario the high energy radiation would always be
energetically significant, thus requiring a careful estimate of the importance 
of photon--photon collisions leading to electron--positron pair production.

\section{The `standard' synchrotron scenario}

Let us briefly summarize the main features of the ISS (for simplicity 
we will also refer to it as the `standard' model).
The emission of the burst, which originates from the conversion of
bulk kinetic energy into random energy, has a duration which is
determined by the central engine.
In particular, in order to generate intermittent and complex
variability patterns, the engine has to produce/eject several
shells of matter.
If these propagate with different Lorentz factors, a faster shell 
will catch up with a slower one and in the interaction a shock will 
develop, which is assumed to be responsible for the acceleration 
of electrons to ultra--relativistic energies.
These would then loose energy by radiating synchrotron photons.

\subsection{Typical radii}

If two shells move with different Lorentz factors, $\Gamma$ and 
$a\Gamma$ (with $a>1$), and are initially separated by $R_{\rm o}$, 
they interact at a distance $R_{\rm i}$ from the central engine, where
\begin{equation}
R_{\rm i}\simeq {2a^2 \over a^2 -1}\, R_{\rm o}\Gamma^2.
\end{equation}
As a reference value in the following we will adopt $a=2$,
corresponding to $R_{\rm i} \simeq (8/3) R_{\rm o}\Gamma^2$.
The initial temporal separation of the two shells, $\sim
R_{\rm o}/c$, also determines the duration of the emission produced by
a single shell--shell collision, as measured by an external observer.

For $a=2$ the final bulk Lorentz factor of both shells is $\sim 1.41 \Gamma$ 
and if they have equal masses the difference between the initial and final 
energy is $\sim$6 per cent of their total kinetic energy.  
In the ISS, this energy is shared among magnetic field, protons and 
electrons.

The other typical distance characterizing the fireball evolution is
the transparency radius $R_{\rm t}$, at which an expanding shell
becomes optically thin to Thomson scattering.
Assuming that each shell carries an energy $E_{\rm s} =10^{50} E_{\rm s,50}$ 
erg in bulk motion, 
$R_{\rm t} \simeq 6\times 10^{12} (E_{\rm s,50}/\Gamma_2)^{1/2}$ cm
\footnote{Here and in the following we parametrize a quantity 
$Q$ as $Q=10^xQ_x$ and adopt cgs units. Primed quantities are 
evaluated in the comoving frame.}.

The standard scenario requires $R_{\rm i} > R_{\rm t}$, i.e. $\Gamma
\gsim 350 (E_{s,50}/R_{0,7}^2)^{1/5}$, in order for the radiation
produced to freely escape (see e.g. Lazzati, Ghisellini \& Celotti 1999).
In the following $E_s$ and $L_s$ stand for the kinetic energy and
power of each shell, respectively, while $L$ stands for the observed
luminosity.

\subsection{The typical synchrotron frequency}

During the shell interaction electrons are instantaneously accelerated
in a collisionless shock, and reach a random Lorentz factor which
corresponds to equipartition with the other forms of energy, i.e.
$\gamma_{\rm eq}=(\Gamma^\prime -1)n_{\rm p} m_{\rm p}/(n_{\rm e} m_{\rm e})$,
where $\Gamma^\prime$ is the Lorentz factor of
one shell in the rest frame of the other, and 
$n_{\rm p}$ and $n_{\rm e}$ are the proton and lepton densities,
respectively.
In the ISS, these are assumed equal (i.e. electron--positron pairs do
not significantly contribute to $n_{\rm e}$).  Deviations from the
equipartition value are parametrized by a dimensionless coefficient
$\epsilon_{\rm e}=\gamma/\gamma_{\rm eq}$.

The out--flowing plasma is magnetized, and a typical/indicative field
value is estimated by assuming that either a significant fraction of
the power is carried as Poynting flux or the field energy in the
emitting region constitutes some fraction $\epsilon_{\rm B}$ of the
randomized energy.
Both possibilities imply that at the distance where the shells interact
the Poynting flux carries a power $L_{\rm B} \equiv R^2\Gamma^2 B^{\prime 2}
c/2 =\epsilon_{\rm B} L_{\rm s}$, where $L_{\rm s}= 4\pi R^2 \Gamma^2  
n_{\rm p}^\prime m_{\rm p} c^3$ is the kinetic power carried by a single
shell.
This corresponds to 
\begin{equation}
B_{\rm eq}^\prime\, =\, 
\left[ 8\pi \epsilon_{\rm B} n_{\rm p}^\prime m_{\rm p} c^2 \right]^{1/2} \,=\, 
\left( { 2\epsilon_{\rm B} L_{\rm s}\over c}\right)^{1/2}\, {1\over \Gamma R}
\end{equation} 
From the above estimates it follows that the typical observed 
synchrotron frequency is
$\nu_{\rm peak}$ $=$ $2 e / (3 \pi m_{\rm e} c)\gamma^2 B^\prime \Gamma /(1+z)$, 
which at $R_{\rm i}$ gives
\begin{equation}
h\nu_{\rm peak}\, \simeq \, 4 \,{\epsilon_{\rm e}^2 (\Gamma^\prime-1)^2\,
\epsilon_{\rm B}^{1/2} L_{\rm s,52}^{1/2} 
\over R_{\rm i, 13} \, (1+z)}\,\,\, {\rm MeV}.
\end{equation}
As mentioned in the Introduction,
the success of the standard scenario in (simply) predicting a typical
observed frequency in remarkable agreement with observations is
probably the strongest piece of evidence pointing towards the
synchrotron process as responsible for the burst emission.

Note that the `equipartition coefficients',
$\epsilon_{\rm B}$ and $\epsilon_{\rm e}$, must be close to unity 
\footnote{The only estimates of these parameters are inferred from the
adoption of this model for the interpretation of the afterglow
emission. This suggests values of $\epsilon_{\rm B}$ and
$\epsilon_{\rm e}$ substantially smaller than unity.} for the
observed value of $\nu_{\rm peak}$ to be recovered
(note that $(\Gamma^\prime -1)$ is of order unity in the ISS).
In turn this also implies/requires that electron--positron pairs
cannot significantly contribute to the lepton density.

The predicted synchrotron spectrum, produced by a quasi
mono--energetic particle distribution, has a flux density
$F_\nu\propto \nu^{1/3}$ up to the cutoff frequency $\nu_{\rm peak}$.
The average observed spectra in the hard X--ray band, which we stress
are typically derived from $\sim 1$ s integrated fluxes, are not 
inconsistent with this shape.  
Nevertheless exceptions exist, including spectra much flatter than
$\nu^{1/3}$, which have already cast some doubts on the synchrotron
scenario (Preece et al. 1998; Lloyd \& Petrosian 1999).

In the following we point out that, just because the integration
and the dynamical timescales are much longer than 
the particle cooling timescales, {\it
the expected synchrotron spectrum in the entire X-- and soft 
$\gamma$--ray band should have a slope $F_{\nu}\propto \nu^{-1/2}$}.
This dramatically exacerbates the discrepancy between the predictions
of the standard scenario and observations.

\section{Radiative cooling time and time integrated spectrum}

Consider the radiative cooling timescale (in the observer frame) of
typical particles emitting synchrotron (and self--Compton) radiation
within the frame of the ISS:
\begin{eqnarray}
t_{\rm cool}\, 
&= &\, {\gamma\over \dot\gamma}\,\, {1+z \over \Gamma}
\,=\, {6\pi m_{\rm e} c (1+z) \over \sigma_{\rm T} B^2 \Gamma \gamma
(1+U_{\rm r}/U_{\rm B})} \, \nonumber \\ 
\, &=& \, 1.14 \times 10^{-7} {\epsilon_{\rm e}^3 (\Gamma^\prime-1)^3 \Gamma_2 
\over 
\nu_{\rm MeV}^2 (1+U_{\rm r}/U_{\rm B})(1+z)}\, \, {\rm s},
\end{eqnarray}
where $U_{\rm r}$ and $U_{\rm B}$ represent the radiation and magnetic
energy densities, respectively.
As already mentioned, the shortest integration times are of the order
of 1 s: this implies that the observed spectrum is produced by a
cooling particle distribution.
Note also that the cooling timescale is much shorter than the dynamical
one, resulting in a relatively efficient radiative dissipation: in
this situation adiabatic energy losses are therefore negligible 
(Cohen et al. 1997; see below).
In particular, 
after one dynamical time $t_{\rm d}= 10^{-2} t_{\rm d,-2}$ s,
the 
cooling electrons emit at the (cooling) observed frequency
$\nu_{\rm cool}\sim 2.2\times 10^{14} (1+z) t_{\rm d,-2}^{-2} \Gamma_2^{-1}
B_4^{-3} (1+U_{\rm r}/U_{\rm B})^{-2}$~Hz, independent 
of $\Gamma^\prime$.
Invoking smaller values of the magnetic field to slow down the cooling 
and obtain $\nu_{\rm cool}$ of the order of few hundreds keV does not help,
since in this case the self--Compton emission dominates over the synchrotron
one (see below).

Since $t_{\rm cool}\propto 1/\gamma$, in order to conserve the particle
number, the instantaneous cooling distribution has to satisfy 
$N(\gamma, t) \propto 1/\gamma$.
When integrated over time, the contribution from particles with
different Lorentz factors is `weighted' by their cooling timescale 
$\propto 1/\gamma$.
Therefore the predicted (time integrated) flux spectrum 
between $\nu_{\rm cool}$ and $\nu_{\rm peak}$ is (e.g. Piran 1999)
\begin{equation}
F_\nu \, \propto \, t_{\rm cool} \, N(\gamma)\,\dot\gamma\, 
{d\gamma\over d \nu}
\, \propto \, 
{\gamma \over \dot \gamma} \, {1\over \gamma}\,  \dot \gamma \, \nu^{-1/2}
 \, \propto \, \nu^{-1/2},
\end{equation}
extending from $\sim h\nu_{\rm cool}\sim$eV to 
$h\nu_{\rm peak}\sim$MeV energies.
We thus conclude that, within the assumptions of the ISS, a major
problem arises in interpreting the observed spectra as synchrotron
radiation.
Let us consider in turn alternative hypotheses, within the same
general frame, which might ameliorate this difficulty.

\subsection{Deviations from equipartition?} 

If one maintains the requirement of observing synchrotron photons at
$\sim$MeV energies, the radiative cooling timescale is almost
independent of $\epsilon_{\rm B}$, i.e. of the assumed value of the
magnetic field (the only dependence being through the ratio
$U_{\rm r}/U_{\rm B}$;  eq.~4).
Furthermore -- again from eq~(4) -- $t_{\rm cool}\sim t_{\rm d}$  requires  
a value of $\epsilon_{\rm e}$ close to $\sim 40$, thus
violating energy conservation (as the electrons would have more energy
than the available one).

On the other hand, if the condition that $\nu_{\rm peak}\sim$1 MeV is
produced by synchrotron is relaxed, the magnetic field intensity can
be smaller than the equipartition value with a consequently longer
synchrotron cooling timescale.
However, as the radiation energy density has to be of the order of
$U^\prime_{\rm r} \sim L_{\rm s}/(4\pi R^2 \Gamma^2 c)\sim 
U^\prime_{\rm B,eq}$ to account for the observed fluxes, 
the inverse Compton cooling would be in any case extremely efficient, 
leading again to short cooling timescales.
Therefore even if the observed radiation is produced by self--Compton
emission of relativistic particles, we face the same problem of
cooling timescales being so short that the spectrum would be steep, 
as discussed below.

\subsection{Particle re-acceleration?}

A further possibility to escape the above conclusion is to assume that
particles are continuously re--heated, thus avoiding the formation of
a cooled particle distribution.
However in the standard ISS new particles are continuously swept
by the shock and are all accelerated to the equipartition energy.
This is a crucial assumption in order to produce a typical observed
peak frequency around a few hundred keV.  
It is thus not possible -- in this scenario -- to continuously
re--accelerate the very same particles, as the energy required would
exceed the available one.
 
Alternatively, relaxing the requirement of the standard scenario, one
can envisage a situation in which only `selected' particles are
steadily accelerated for the entire duration of the shell--shell
interaction.
In this case an extreme fine (and unlikely) tuning is required: in fact, 
to be consistent with the total energetics, the selected particles have to: 
i) be fixed in number (only a fraction 
$\sim t^\prime_{\rm cool}/(\Delta R^\prime /c)$
of the total number of particles can be accelerated);
ii) be always the same; 
iii) achieve $\gamma\sim\gamma_{\rm eq}$ even in the absence
of an equipartition argument.

It would be also plausible to assume that the emission is produced by a 
power--law distribution of electrons resulting from continuous acceleration 
and cooling.
Indeed, for an energy distribution $\propto \gamma^{-p}$ (with
$p>0$) only a minority of particles attain the maximum energy.
But -- besides having to keep all the particles accelerated for the
entire duration of the shell--shell interaction -- the relative number 
of the most energetic particles requires $p >2$, leading to
a spectrum even steeper than $\nu^{-1/2}$.

We therefore conclude that re--acceleration does not avoid the spectral 
discrepancy, even when relaxing some of the key assumptions of the standard
scenario.

\subsection{Strongly varying magnetic field?}

Let us consider the case in which the magnetic field attains a value
close to the equipartition one only in a very limited region (e.g. near
the shock front), while is weaker elsewhere.
In this situation the synchrotron cooling is mostly effective within
this region only and the particles may not have time to significantly cool.
Therefore in principle a synchrotron spectrum $\propto \nu^{1/3}$
might be produced.
This requires that particles loose much less than half of
their energy in the radiative zone, since even a reduction of a factor
two of their Lorentz factor would imply that a spectrum $\nu^{-1/2}$ is
produced in a range spanning a factor four in frequency 
(this may correspond to the entire BATSE energy range).

Therefore it would be required that:

\noindent
i) the synchrotron process in the most radiative region $must$
be inefficient, since it has to reduce the electron energy at most by
a small fraction;

\noindent
ii) away from this zone, particles continue to rapidly cool by
self--Compton and - at a reduced rate - by synchrotron emission. 
The inverse Compton process then becomes the dominant cooling
mechanism.

\noindent
iii) since the cooling is very rapid anyway, the self--Compton
emission itself would produce a time integrated (over $t_{\rm d}$) steep
spectrum.

We conclude that the net effect of having a strong magnetic field
confined in a limited region is to decrease the total synchrotron power
in favor of the self--Compton one, whose spectrum would in any case be
steep (see below).

\subsection{Adiabatic losses?}

Suppose that particles are accelerated in compact regions that rapidly expand
because of internal pressure.  
Adiabatic losses dominate particle cooling as soon as the particle
Lorentz factor decreases below some critical $\gamma_{\rm ad}$, thus
generating a spectrum $\propto \nu^{1/3}$ below the synchrotron
frequency $\nu_{\rm ad}$ (corresponding to $\gamma_{\rm ad}$), and
steeper above (Cohen et al. 1997).
However this possibility faces two severe problems, both related to
the overall efficiency, being required that:

\noindent
i) each electron loses only a small fraction of its energy radiatively
(i.e. $\gamma_{\rm ad}/\gamma$ must be greater than $\sim 1/2$);

\noindent
ii) the emitting regions are very compact, for adiabatic losses
to be significant.
This implies that the transformation of bulk into random energy
does not occur in a shell subtending the entire ejection solid angle.
Photon and electron densities have then to be higher to account for
the observed luminosity, thus enhancing the inverse Compton process.

\section{Importance of the relativistic Inverse Compton process}

As long as the scattering optical depth $\tau_{\rm T}$ of the electron
in the emitting region is smaller than unity, the importance of the
relativistic inverse Compton process with respect to the synchrotron
one is measured by the (relativistic) Compton parameter $y^\prime\equiv
\tau_{\rm T} \gamma^2\beta^2  = \sigma_{\rm T}\gamma^2\beta^2
n^\prime_{\rm e} c t^\prime_{\rm cool}$.
The width of the region corresponds to a cooling length, $c
t^\prime_{\rm cool}$, as assumed within the standard ISS.

Since the magnetic field intensity is related to the proton density
$n^\prime_{\rm p}$ we obtain

\begin{equation}
y^\prime\, =\, {3 \over 4}\, {\epsilon_{\rm e}\over \epsilon_{\rm B}}\, 
{n^\prime_{\rm e} \over n^\prime_{\rm p} }\,
{\Gamma^\prime -1 \over 1+U^\prime_{\rm r} /U^\prime_{\rm B}}.
\end{equation}

This implies that the inverse self--Compton power is of the same order of
the synchrotron one \footnote{This is true as long as the scattering
process is in the Thomson regime, i.e. $\gamma < 800 B_4^{\prime-1/3}$ for
the first order Compton scattering.}, and is emitted 
at a typical observed energy (for the first order)
\begin{equation}
h\nu_{\rm c} \simeq \, \gamma^2 h\nu_{\rm s}\, \simeq \, 13\, 
{ \epsilon_{\rm e}^4 (\Gamma^\prime -1)^4 
\epsilon_{\rm B}^{1/2} L_{\rm s,52}^{1/2}
\over R_{\rm i, 13} (1+z)}\,\, {\rm TeV}.
\end{equation}
Two points are worth being stressed. 
First, the strong dependence of $\nu_{\rm c}$ on the equipartition
parameter $\epsilon_{\rm e}$.
Furthermore, although in the comoving frame this typical Compton 
frequency is a factor $\Gamma$ lower, it can still largely exceed 
the pair production threshold (see below).

It has been mentioned and implicitly assumed above that even in the
hypothesis that the hard X--ray burst radiation is due to
self--Compton emission, the argument of the fast cooling producing a
steep spectrum applies. 
Let us consider  this possibility more closely, and in particular 
the first order inverse Compton spectrum.

Although in this case the typical electron energies required are
smaller, the cooling timescales are still much shorter than the 
dynamical time: in fact, to produce $\sim$MeV photons by
the first order Compton scattering, $\gamma\sim 83 [\nu_{\rm MeV}
(1+z)/(\Gamma_2 B'_4)]^{1/4}$ with a corresponding cooling time
\begin{equation}
t_{\rm cool} = 1.4\times 10^{-5} { R_{\rm i,13}^2\Gamma_2^{5/4} 
B_4^{\prime~1/4}  
(1+z)^{3/4} \over 
L_{50} (1+U_{\rm B}/U_{\rm r}) \nu_{\rm MeV}^{1/4} } 
\,\, {\rm s}.
\end{equation}
Here $L=10^{50}L_{50}$ erg s$^{-1}$ is the (observed) luminosity produced 
by a single shell.

Furthermore, by following the same arguments leading to eq.~(5), the
predicted time--integrated spectrum results $F_\nu \propto \nu^{-3/4}$,
i.e. {\it even steeper} than $\nu^{-1/2}$.

A further difficulty of interpreting the burst emission as first order
scattering is that if the inverse Compton power exceeds the
synchrotron one by a certain factor, then each higher Compton order
will dominate over the previous one by the same amount, until the
typical emitted frequency reaches the electron energy.
Only a small fraction of the radiated power would therefore be
observed (in the hard X--ray band).

\section{Pair production}

The above results indicate that the time integrated spectrum predicted
by the standard scenario is steeper than observed.
Furthermore, the power emitted through the self--Compton process
should be comparable to -- if not more than --  the synchrotron one, 
and emitted at energies exceeding the pair production threshold. 
It is thus compelling to estimate the importance of photon--photon
collisions producing electron--positron pairs.

Setting $x\equiv h\nu/(m_{\rm e} c^2)$, the energy threshold for
photons of energy $x$ is $x_{\rm T} = 2/[x(1-\cos\theta)]$, where
$\theta$ is angle between the two photon directions. 
Also, the photon--photon collision rate is proportional to 
$(1-\cos\theta)$.

The result of the integration of the photon--photon cross section over
the energy of the target photons can be well approximated by
$(\sigma_{\rm T}/5) x_{\rm T} n_\gamma(x_{\rm T})$ (Svensson 1987),
where $n_\gamma(x_{\rm T})$ is the number density of photons of energy
$x_{\rm T}$, which is related to the observed luminosity 
$L(x_{\rm T})$ by 
$x_{\rm T} n_\gamma(x_{\rm T})= L(x_{\rm T})/(4\pi m_{\rm e} c^3 R^2)$.

The optical depth for pair production in the observer frame can be
then expressed as
\begin{equation}
\tau_{\gamma\gamma}(x)\, =\,
{\sigma_{\rm T} \over 20\pi} \, { L(x_{\rm T}) 
\langle 1-\cos\theta\rangle \, \over R m_{\rm e} c^3} {\Delta R\over R}.
\end{equation}
$\Delta R$ may represent the width of the emitting shell or,
alternatively, the typical scale over which the emitted photons might
interact (i.e. $\Delta R\sim R$), depending whether we are interested in 
the pair production within the shell or also outside it.

Since the source is moving relativistically, all photons appear to be
emitted quasi--radially, and interact with a typical angle 
$\sin\theta\sim 1/\Gamma$ (corresponding to $\cos\theta\sim \beta$), 
and thus $\langle 1-\cos\theta\rangle\sim 1/\Gamma^2$.  
If the typical size of the fireball is estimated by time variability,
$R\sim c t_{\rm var}\Gamma^2$, we have:
\begin{equation}
\tau_{\gamma\gamma}(x)\, =\, {\sigma_{\rm T} \over 20\pi} \, 
{L(x_{\rm T}) \over t_{\rm var} m_{\rm e} c^4 \Gamma^4 } \,
{\Delta R\over R}
\end{equation}
In this form $\tau_{\gamma\gamma}(x)$ can be estimated even without
detailed spectral information.
If the observed spectrum is a power law $L(x) \propto
x^{-\alpha}$, with $\alpha<1$ up to a  maximum energy $x_{\rm max}$, the
observed luminosity at threshold is related to the total luminosity
$L$ by $L(x_{\rm T}) = [(1-\alpha)/2^\alpha](L/x_{\rm max}^{1-\alpha})
(x/\Gamma^2){^\alpha}$, giving
\begin{eqnarray}
\tau_{\gamma\gamma}(x)\, 
&=& \, {(x/2)^\alpha   \over \Gamma^{4+2\alpha}}\,
{(1-\alpha) \ell \over 20\pi  x_{\rm max}^{1-\alpha}}
\, {\Delta R\over R}, 
\end{eqnarray}
where the compactness $\ell \equiv \sigma_{\rm T} L / (t_{\rm var}
m_{\rm e} c^4)$ has been introduced.  
Note that the optical depth increases with photon energy $x$.

For illustration, consider a burst with $\Gamma=10^2$ lasting 
$t_{\rm var}=10$ ms.
1 GeV photons ($x=2000$) mostly interact with target photons of energies
$x_{\rm T}=10~\Gamma_2^2$, i.e. $\sim$ 5 MeV.
Assume that the observed luminosity at these energies is
$L(x_{\rm T})=10^{50}$ erg s$^{-1}$. From eq.~(10) we have
\begin{equation}
\tau_{\gamma\gamma}(x=2000)\, \sim 1.4\times 10^3 \, {L_{50}(x_{\rm T}) 
\over \Gamma_2^4 } \, {\Delta R\over R}.
\end{equation}
Since the optical depth is so large, all the high ($\sim$GeV)
energy emission can be easily absorbed, unless $\Gamma>10^3$.

For $\Gamma<10^3$ relativistic pairs can then be copiously produced.
They will immediately cool radiatively, initiating a pair cascade, strongly
affecting the primary spectrum, and possibly even the dynamics.
(Equilibrium) spectra produced by pair cascades have been extensively
studied in the past in the context of nuclear AGN emission.
The general outcome is that pairs act as reprocessors of the high
energy emission, which is absorbed and ultimately reprocessed into lower
energy radiation.
This corresponds to a steepening of the spectrum, thus exacerbating
the discrepancy with the observed bursts.

Furthermore, the lepton density $n_{\rm e}$ may become substantially 
larger than the proton one, $n_{\rm p}$, and thus the equipartition 
energy $\gamma_{\rm eq}$ smaller than $m_{\rm p}/m_{\rm e}$.

\section{Discussion}

The main point stressed in this paper is the inadequacy of the
synchrotron and inverse Compton emissions from ultra-relativistic
electrons to account for the observed burst spectra -- at least within
the scenario invoking internal shocks for the dissipation of the
fireball bulk kinetic energy.

This is independent of detailed assumptions and directly follows from
the extremely short cooling timescales -- compared to the dynamical
time which is in turn much smaller than the integration
one -- {\it required} by the ISS, which lead to a steep emitted spectrum. 
No alternative hypothesis, which could alleviate this spectral discrepancy, 
has been found.
Furthermore it is stressed that within the ISS scenario electron--positron 
pairs would be naturally and copiously produced, contrary to the basic 
model assumptions.

The situation is somewhat paradoxically: the emission mechanism at the
origin of the burst radiation must be very efficient, and the synchrotron 
and inverse Compton mechanisms by relativistic particles are indeed very 
efficient radiation processes, but just because of the very rapid cooling 
their predicted spectrum is too steep.

One is therefore forced to look for alternatives.
In the dense photon environment of the internal shock scenario
a highly efficient viable alternative radiation mechanism may be
Comptonization by a quasi--thermal particle distribution, as proposed
by Ghisellini \& Celotti (1999) (see also Thompson 1994; Liang 1997;
Stern 1999; Liang et al. 1999).
In this model, the conversion of bulk kinetic into random energy may
still be due to shell--shell collisions, but the typical energy of the
radiating particle is sub--relativistic, being fixed by the balance
between the acceleration and the cooling processes.
There is still equipartition between magnetic field, leptons and
protons energies, but in a time integrated sense: all leptons are
accelerated up to small energies, but for the entire duration of the
shell--shell collision.
Electron--positron pairs can be produced, and may even be the key
ingredient to lock the particle energies in the observed range.
These sub- or mildly relativistic particles would then emit
self--absorbed cyclo--synchrotron photons and a Comptonization
spectrum with a typical slope $F_\nu\propto \nu^0$ plus a Wien peak
located where photon and particle energies are equal.
The predicted spectral and temporal evolutions from this model are 
under investigation.

\section*{Acknowledgments}
AC acknowledges the Italian MURST for financial support. This research
was supported in part by the National Science Foundation under Grant
No. PHY94-07194 (AC).

\end{document}